\RequirePackage{lineno}
\documentclass[aps,prb,showpacs, groupedaddress,twocolumn,superscriptaddress]{revtex4}
\usepackage{amsmath,graphicx,latexsym,times,color}
\usepackage{setspace}
\usepackage{hyperref}
\usepackage{array}
\usepackage{titlesec}
\usepackage{physics}

\begin{document}

\title{Symmetry aspects of spin-filtering in molecular junctions: hybridization and quantum interference effects}

\author{Dongzhe Li}
\email{dongzhe.li@uni-konstanz.de}
\affiliation{Department of Physics, University of Konstanz, 78457 Konstanz, Germany}

\author{Rajdeep Banerjee}
\affiliation{Theoretical Sciences Unit and School of Materials, Jawaharlal Nehru Centre for Advanced Scientific Research, Jakkur, Bangalore 560064, India}

\author{Sourav Mondal}
\affiliation{Theoretical Sciences Unit and School of Materials, Jawaharlal Nehru Centre for Advanced Scientific Research, Jakkur, Bangalore 560064, India}

\author{Ivan Maliyov}
\affiliation{DEN, Service de Recherches de M\'etallurgie Physique, CEA, Universit\'e Paris-Saclay, 91191 Gif-sur-Yvette Cedex, France}

\author{Mariya Romanova}
\affiliation{Laboratoire des Solides Irradi\'es, Ecole Polytechnique, CEA-DRF-IRAMIS, CNRS UMR 7642, 91120 Palaiseau, France}

\author{Yannick J. Dappe}
\affiliation{SPEC, CEA, CNRS, Universit\'e Paris-Saclay, CEA Saclay, 91191 Gif-sur-Yvette Cedex, France}

\author{Alexander Smogunov}
\affiliation{SPEC, CEA, CNRS, Universit\'e Paris-Saclay, CEA Saclay, 91191 Gif-sur-Yvette Cedex, France}

\date{\today}

\begin{abstract}
Control and manipulation of electric current and, especially, its degree of spin polarization (spin filtering) across single molecules are currently of great interest in the field of molecular spintronics. We explore one possible strategy based on the modification of nanojunction symmetry which can be realized, for example, by a mechanical strain. Such modification can activate new molecular orbitals which were inactive before due to their orbital mismatch with the electrode’s conduction states. This can result in several important consequences such as (i) quantum interference effects appearing as Fano-like features in electron transmission and (ii) the
change in molecular level hybridization with the electrode’s states. We argue that the symmetry change can affect very differently two majority- and minority-spin conductances and thus alter significantly the resulting spin-filtering ratio as the junction symmetry is modified.We illustrate the idea for two basic molecular junctions: Ni/benzene/Ni (perpendicular vs tilted orientations) and Ni/Si chain/Ni (zigzag vs linear chains). In both cases, one highest occupied molecular orbital (HOMO) and one lowest unoccupied molecular orbital (LUMO) (out of HOMO and LUMO doublets) are important. In particular, their destructive interference with other orbitals leads to dramatic suppression of majority-spin conductance in low-symmetry configurations. For a minority-spin channel, on the contrary, the conductance is strongly enhanced when the symmetry is lowered due to an increase in hybridization strength. We believe that our results may offer a potential route for creating molecular devices with a large on-off ratio of spin polarization via quantum interference effects.
    
\end{abstract}

\renewcommand{\vec}[1]{\mathbf{#1}}

\maketitle

\section{ Introduction }
\label{Intro}

Theoretical understanding and experimental realization of electron transport through single-molecule junctions with ferromagnetic electrodes,  especially with the goal to control and manipule its degree of spin-polarization, is currently of great importance in the field of molecular spintronics \cite{Rocha-2005, Sanvito-2010} -- a rapidly developing branch of nanoelectronics. Among many properties of interest is the spin-filtering ratio\cite{comment2} (or spin-polarization SP), which measures the degree of spin-polarization of the electronic conductance.High values of SP are usually accompanied by large magneto-resistance ratios (MRR) 
measuring the difference in conductance between parallel and antiparallel magnetic orientations of two ferromagnetic electrodes.
Is  is  also a great challenge to suggest possible systems and measurements where solid symmetry arguments and quantum effects such as destructive (or constructive) interference could play an important role in spin-polarization and could be tested experimentally.

At the single molecule scale, more than 100\% of MRR were predicted theoretically for Fe/C$_{60}$/Fe magnetic junctions \cite{C60-Fe}, and moderate values of MRR up to 60\% were recently reported \cite{Schmaus-2011, Bagrets-2012}. These effects were attributed to the spin dependent hybridization occurring at the ferromagnet/molecule interface. Also, large spin-polarizations of different organic molecules due to hybridization with magnetic substrates were reported \cite{Caffrey-2013, Atodiresei-2010, Kawahara-2012, Dongzhe-C60-2016}. This can result in large tunneling magnetoresistance (TMR) as was shown, for example, for C$_{60}$ molecules on Cr(001) terraces \cite{Kawahara-2012}. Regarding atomic nanocontacts, giant magnetoresistance (GMR) values of about 70\% were reported recently for Co/Au/Co metallic junctions \cite{Sivkov-2014} as a result of strong perturbation of $s$-states at Au contact atoms, and very large spin-polarizations of the current were found in half-metallic NiO monatomic junctions \cite{Jacob-2006}. Rather detailed analysis done by Bagrets and co-workers \cite{Bagrets-2004} for short (3-atom long) chains of various species joining two Co electrodes has also revealed, in some cases, MRR as large as 50\%. More recently, enhanced spin injection has been reported in amine-ended molecular junction via mechanical strain which was explained by pronounced spin up transmission due to hybridization between $d_{yz}$ and $\pi$-orbitals of a molecule \cite{Tang-2016}.

\begin{figure*}[t]
	\centering
	\includegraphics[width=1.0\linewidth]{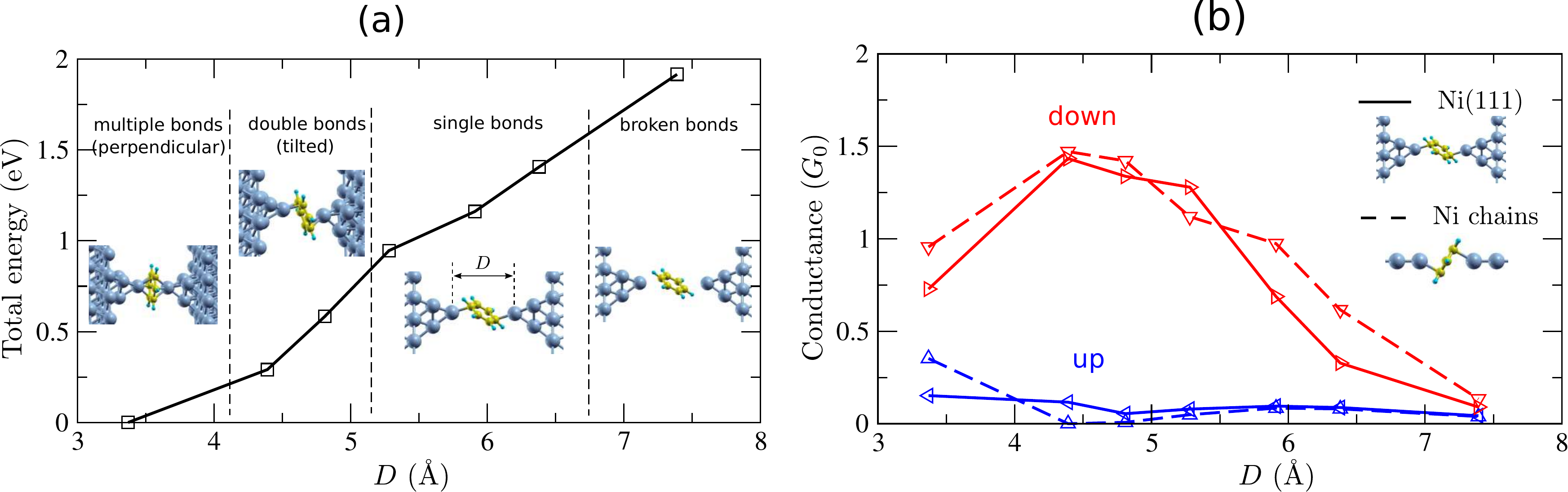}
	\caption{
		DFT calculations for Ni/Benzene/Ni junctions: 
		(a) total energy as a function of electrode-electrode separation $D$ (first point was set as zero). 
		The insets show the representative relaxed geometries at different values of $D$;
		(b) spin-dependent conductance (given by the Fermi energy transmission) as a function of stretching 
		for both crystalline Ni(111) electrodes (solid lines) and  model semi-infinite Ni chains (dashed lines).
	}
	\label{stretching-E-G}
\end{figure*}

We have recently suggested \cite{Smogunov-2015, Dongzhe-2016} a mechanism, based on robust symmetry considerations, which can lead to perfect spin-filtering and to huge (ideally, infinite) MRR ratios across some specific molecules joining ferromagnetic electrodes. Due to symmetry match/mismatch between molecular orbitals (involved in electron transport) and ferromagnetic electrodes, the charge transport was carried by only few (spin down) $d$-states while all the $s$-channels were fully blocked at the metal-molecule interface. Later on, some indications on highly spin-polarized conductance in Ni/O/Ni junctions have been experimentally reported by the group of Oren Tal\cite{Vardimon-2015} supporting our symmetry arguments.

In the present paper, we develop further our symmetry arguments with the aim to suggest a possible way for tuning spin-filtering ratio in molecular junctions. In particular, a mechanical strain is proposed to modify the junction symmetry which could switch on or off new 
conduction channels. The idea is illustrated on two possible examples: a benzene molecule (perpendicular vs tilted configuration) and a short Si chain (zigzag vs linear configuration) connecting two ferromagnetic Ni electrodes. We find, in particular, that in both cases the majority spin conductance is strongly quenched in low symmetry configurations (a tilted benzene or a zigzag Si chain) which is rationalized as a result of destructive quantum interference. The minority spin conductance is on the contrary significantly enhanced due to stronger hybridization of molecular orbitals with ferromagnetic electrodes.

It should be noted that previous studies on quantum interference in electron transport focused on different 
geometrical conformations or connections of molecules to nonmagnetic electrodes \cite{Hansen2009, Solomon-2008, Aradhya2012, Panu-2017, Borges-2017}. The mostly discussed effect is, for example, the destructive quantum interference occurring when the benzene molecule is connected to electrodes in meta or ortho configuration contrary to the para orientation \cite{Hansen2009, Solomon-2008,Miao-2018}. 
Very recently, the destructive $\sigma$-inteference was also reported in nonmagnetic molecular junctions, resulting in a complete suppression of the transmission close to the Fermi level \cite{Garner-2018}. 
To the best of our knowledge, our results demonstrate for the first time how the quantum interference can be 
exploited for strong polarization of electric current in molecular junctions with ferromagnetic electrodes.
We show moreover that the destructive interference (blocking fully one spin channel) can be easily tuned (switched on or off) 
by mechanical strain modifying the symmetry of the junction.

\begin{figure*}[t]
	\centering
	\includegraphics[width=1.0\linewidth]{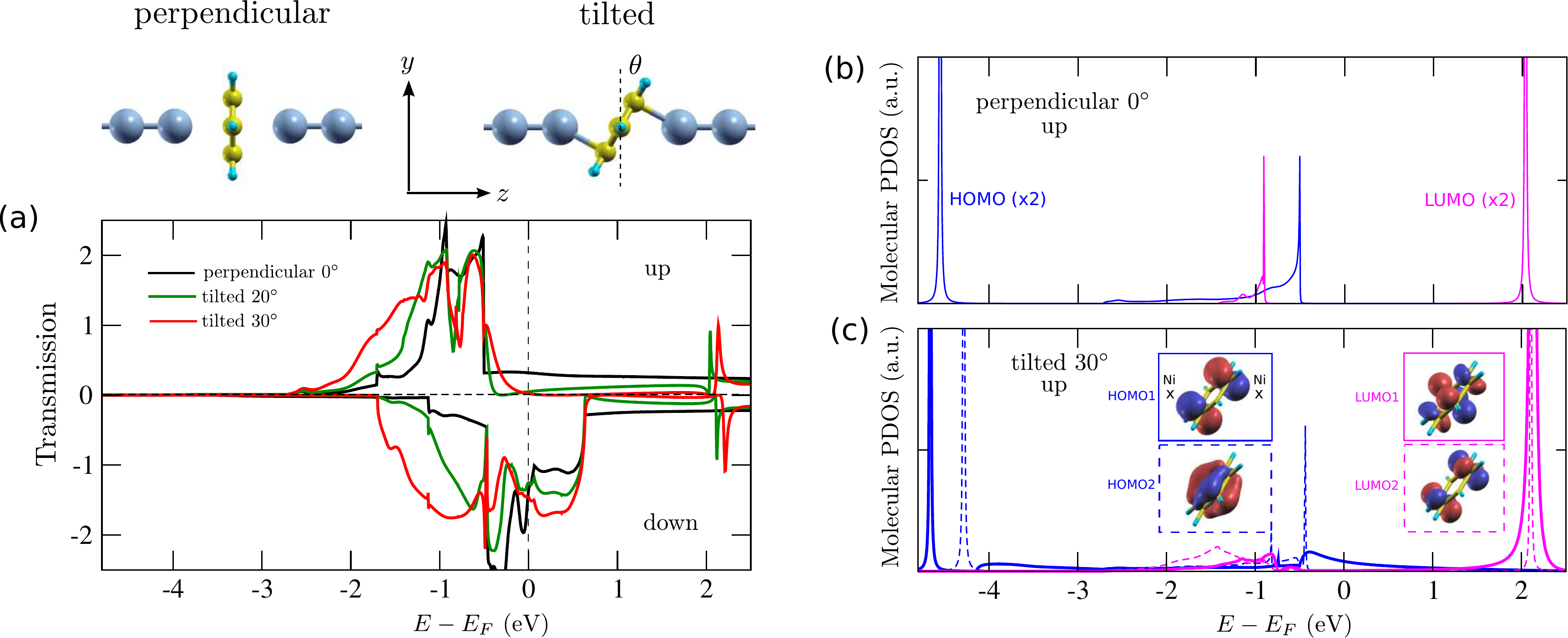}
	\caption{
		TB transport calculations of model Benzene junctions with Ni atomic chains as electrodes
		for different Benzene tilting angle $\theta$:
		(a) spin-resolved transmissions as a function of $\theta$. 
		Due to destructive interference from HOMO1 and LUMO1 orbitals, activated by tilting, 
		the spin up transmission gets fully suppressed at the Fermi energy; 
		(b) DOS projected onto HOMO and LUMO molecular orbitals for spin up channel in perpendicular configuration; 
		(c) same for tilted configuration of $\theta=30^{\circ}$. 
		Wavefunctions of all the HOMO and LUMO orbitals are shown on insets where 
		the positions of contact Ni chain atoms are also indicated. 
		Only HOMO1 (solid blue) and LUMO1 (solid pink) orbitals, due to their symmetry (even with respect to the $YZ$ plane), 
		have nonzero overlap with Ni $s$-channel.
	}
	\label{Ni1Dchain}
\end{figure*}

\section{Results and discussion}
The spin-polarized electronic structure calculations were performed using Quantum-ESPRESSO (QE) \cite{Giannozzi2009} package based on the density functional theory (DFT). We used the PBE \cite{pbe-1996} exchange-correlation functional with ultrasoft pseudopotentials to describe electron-ion interactions. For electron transport two codes have been used. 
Spin-polarized conductances (given by electron transmission at the Fermi energy) for different molecular junctions shown in Fig.~\ref{stretching-E-G} 
were evaluated using the plane-wave scattering based approach as implemented in PWCOND code \cite{alex-2004}.
All the energy dependent transmissions presented on other figures were calculated using home-made tight-binding (TB) code 
\cite{Gabriel-2008} with TB parameters extracted from \textit{ab initio} QE calculations.  
The standard non-equilibrium Green's function (NEGF) formalism was employed to obtain the transmission function:  
\begin{equation}
T(E)={Tr}[\Gamma_L(E)G^r(E)\Gamma_R(E)G^a(E)],
\end{equation}
where $G^{r,a}$ are the retarded and advanced Green's functions of the central region (including the molecule and some portions 
of left and right electrodes), and $\Gamma_{L,R}$ are the coupling matrices of the central region to the left and right electrodes,
respectively. For spin-polarized calculations all the quantities depend also on the spin index but two spin channels are not coupled
(without spin-orbit interactions and for collinear magnetism).

\subsection{Strain-dependent spin filtering in Ni/benzene junctions}

We start our discussion with Ni/benzene/Ni molecular junctions, shown in Fig. \ref{stretching-E-G}, 
which were recently studied extensively in respect to large anisotropic magnetoresistance\cite{orenAMR}.   
To determine possible geometries we stretch gradually the junctions up to the breaking point, starting from the benzene molecule which is perpendicular to the direction of the current flow. The molecular contact is described in a supercell containing the benzene molecule and 
two 4-atom Ni pyramids attached to a Ni(111) slab containing five and four atomic layers on the left and right side, respectively, see insets of Fig. \ref{stretching-E-G}. During the stretching process, we move apart the two electrodes stepwise and relaxe all the atomic positions except 
of two bottom Ni layers on both sides (which were kept at their bulk-like positions) until the junction was broken.

As presented in Fig. \ref{stretching-E-G} (a), our \textit{ab initio} structural optimization has shown that upon increasing the electrode-electrode 
separation $D$ the benzene molecule is connected to Ni electrodes in three different ways before the breaking point: 
(i) by all six C atoms bound symmetrically to Ni apex atoms -- direct $\pi$-bound perpendicular orientation for small separations,
(ii) by two C atoms on each side -- mixed $\pi$- and $\sigma$-bound tilted orientation for intermediate separations, and 
(iii) by only one C atom on each side (larger separations). These results are similar to those reported previously for Pt/benzene/Pt nanojunctions \cite{Kiguchi-2008, Yelin-2013}. Very recently Rakhmilevitch $et~al.$\cite{Rakhmilevitch-2016} successfully created single-molecule junctions based on Ni electrodes and benzene molecules using MCBJ technique, which supports our theoretical results. 
Thus, during the stretching, the symmetry of the molecule connection is lowered from the highly symmetric configuration (i) to low symmetry geometries (ii) and (iii).

The spin-polarized conductance as a function of stretching is shown in Fig. \ref{stretching-E-G}b. The results for two types of electrodes 
are presented: realistic Ni(111) electrodes (solid lines) which were used for atomic relaxations of Fig. \ref{stretching-E-G}a and
model Ni chain electrodes. In both cases the conductance is found to be highly strain-dependent in both spin channels. 
One can distinguish three regimes clearly correlating with three Benzene orientations. 
In perpendicular configuration i) the conductance is rather low and moderately spin-polarized. 
In tilted orientation ii) the spin down conductance is strongly enhanced while the spin up conductance is on the contrary suppressed which gives rise to dramatic increase in spin polarization of conductance. Very interestingly, the suppression of spin up conductance is complete for model Ni electrodes -- the point which will be addressed in detail in the following. 
In single bond regime iii) the spin down/up conductance starts to decrease/increase reducing again the spin-filtering ratio.

\subsection{Ni chain/benzene junctions: Interference effects}

With the aim to understand a dramatic change of spin-polarized conductance between i) and ii) orientations, we consider first a model case of Ni chain 
electrodes, as shown at the top of Fig.~\ref{Ni1Dchain}. We plot in Fig.~\ref{Ni1Dchain} (a) the spin-resolved transmission function for three configurations, namely perpendicular and tilted with $\theta=20^{\circ}$ and $30^{\circ}$ configurations. All the three geometries have the same separation distance of $D$ =  4.39 \AA~ taken from the most representative double bonds configuration (see Fig. \ref{stretching-E-G}). In order to understand better electron transport, we also plot on the middle and lower panels the density of states (DOS) projected onto molecular frontier HOMO and LUMO orbitals (for simplicity for spin up states only). It is known (see, e.g., Ref. \cite{Smogunov-2015}), that the Ni atomic chain has only one $s$ band around the Fermi energy for spin up channel while five more $d$ bands are also available for spin down one.

\begin{figure*}[t]
	\centering
	\includegraphics[width=.8\linewidth]{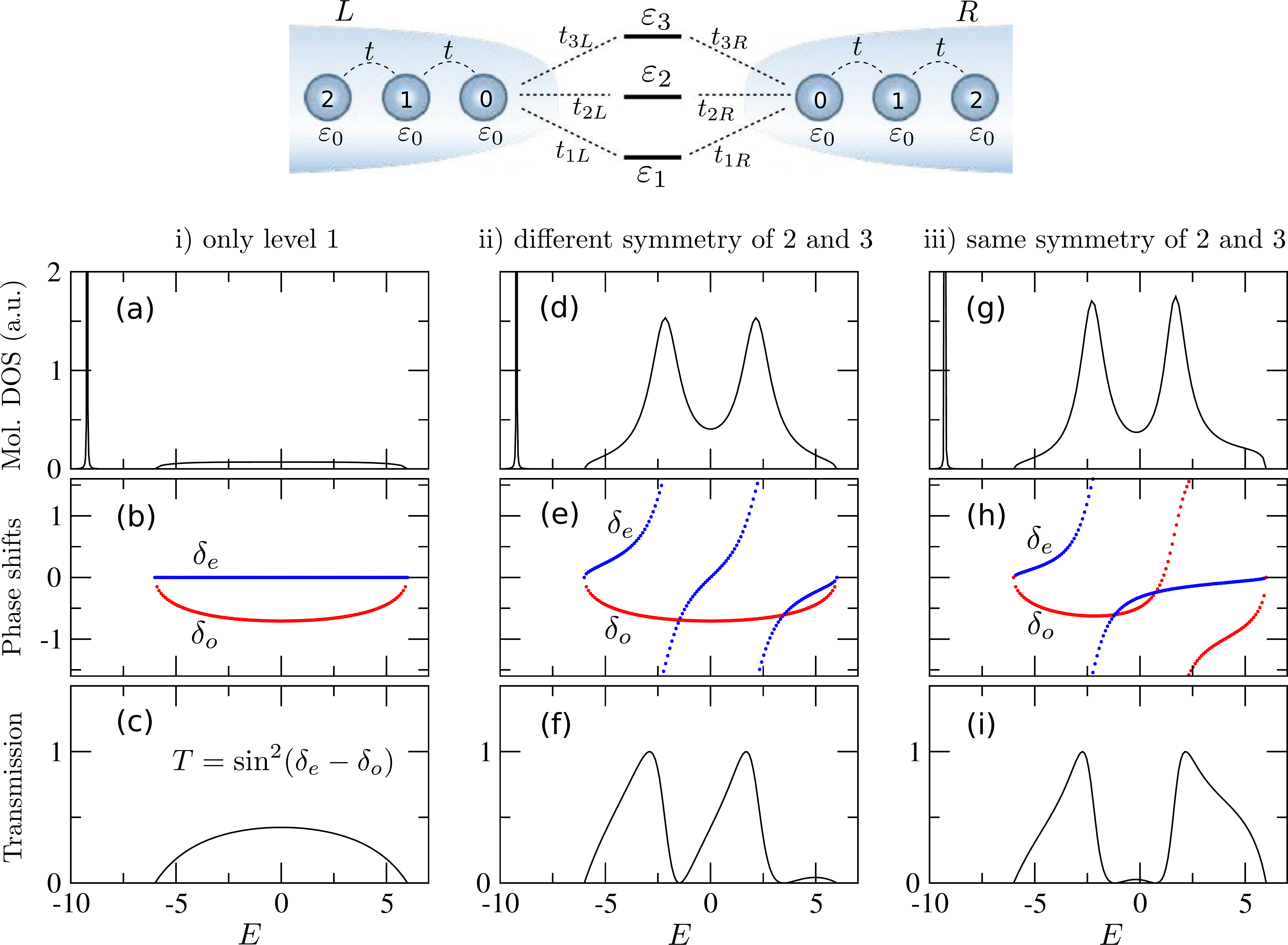}
	\caption{
		Simple 3-level 1D model providing different quantum interference patters in transmission. Molecular (summed over all three levels) DOS,
		even/odd phase shifts and derived from them transmission functions are shown on upper, middle, and lower panels, respectively.  
		Three different cases are considered: 
		i) only first level, $\varepsilon_1=-7$, is coupled to chains with $t_{1L}=-t_{1R}=3$, panels (a-c);
		ii) levels 2 and 3, $\varepsilon_2=-2, \varepsilon_3=2$, are also added, first coupled symmetrically ($t_{2L}=t_{2R}=1$) and the 
		second one - antisymmetrically ($t_{3L}=-t_{3R}=1$), panels (d-f); 
		iii) levels 2 and 3 are coupled both symmetrically, $t_{2L}=t_{2R}=t_{3L}=t_{3R}=1$, panels (g-i).
		Electrode conduction band spreads within the energy interval [-6;6], which corresponds to $\varepsilon_0=0,~t=3$, 
		and the Fermi level is set to zero.
	}
	\label{phaseshifts}
\end{figure*}

We first analyze the spin up channel. Clearly, its conductance (determined by transmission at the Fermi energy) gets significantly suppressed when the molecule is tilted. Interestingly, this effect can be interpreted as a result of destructive quantum interference.
In perpendicular configuration, the frontier orbitals (two-fold degenerate LUMO and HOMO) are all strictly orthogonal to the Ni $s$-channel 
(see insets in \ref{Ni1Dchain}c) and do not therefore manifest in transport at energies from $-0.5$ to $2.7$~eV with respect to the Fermi level, where only $s$-channel is present in Ni chains. Small non-zero transmission here is attributed to the HOMO-2 orbital which is rotationally symmetric (see Appendix A, Fig. \ref{free-bz}) and has thus a perfect overlap with Ni $s$-states providing a rather constant tunneling transmission of about $0.3~G_0$ 
at the Fermi level. For tilted orientation only one symmetry plane $YZ$ is left. Two orbitals, HOMO1 and LUMO1, which are both symmetric with respect to that plane (insets of Fig. \ref{Ni1Dchain}c) can now overlap with the (symmetric) $s$-channel of Ni while two others, HOMO2 and LUMO2, are antisymmetric and therefore remain uncoupled.      
This can be seen from the DOS analysis shown in Fig. \ref{Ni1Dchain}c: 2-fold degenerate HOMO splits into two peaks located at about $-4.7$ eV (HOMO1) and $-4.3$ eV (HOMO2), and the finite DOS at [$-0.4$ eV; $1.0$ eV] of HOMO1 is due to finite hybridization with $s$-states of Ni. 
Similarly, the two LUMO orbitals, located at about $2$ eV, 
show up as  a sharp (delta-like) peak (LUMO2) and a rather broad resonance (LUMO1). 
These additional conduction paths provided by HOMO1 and LUMO1 interfere with the former tunneling one resulting in a Fano-like shape
of spin up transmission and its strong suppression in a broad energy window of [-0.4  eV; 2.0 eV].

For spin down polarization, the conductance is dominated by additional $d$ channels. In particular, two $d_{xz,yz}$ Ni bands appear 
at $E<0.7$~eV increasing significantly the spin down transmission. Those bands have non-zero overlap with HOMO orbitals 
even in perpendicular configuration and hybridize also to LUMO states when the molecule is tilted.
The stronger hybridization of spin down $d_{xz,yz}$ bands to frontier orbitals is therefore responsible for increase of  
transmission at $E<0.7$~eV when the molecule is tilted.

It should be noted here that more sophisticated transport
calculations with PWCOND have produced very similar transmission
curves (see Appendix B, Fig. \ref{sup2}), reproducing very well the same features, in particular, the
destructive interference pattern in the spin-up channel. This validates and supports our TB calculations, which allow us to
explore in more detail our findings and the mechanism behind
them.

\begin{figure*}[t]
	\centering
	\includegraphics[width=1.0\linewidth]{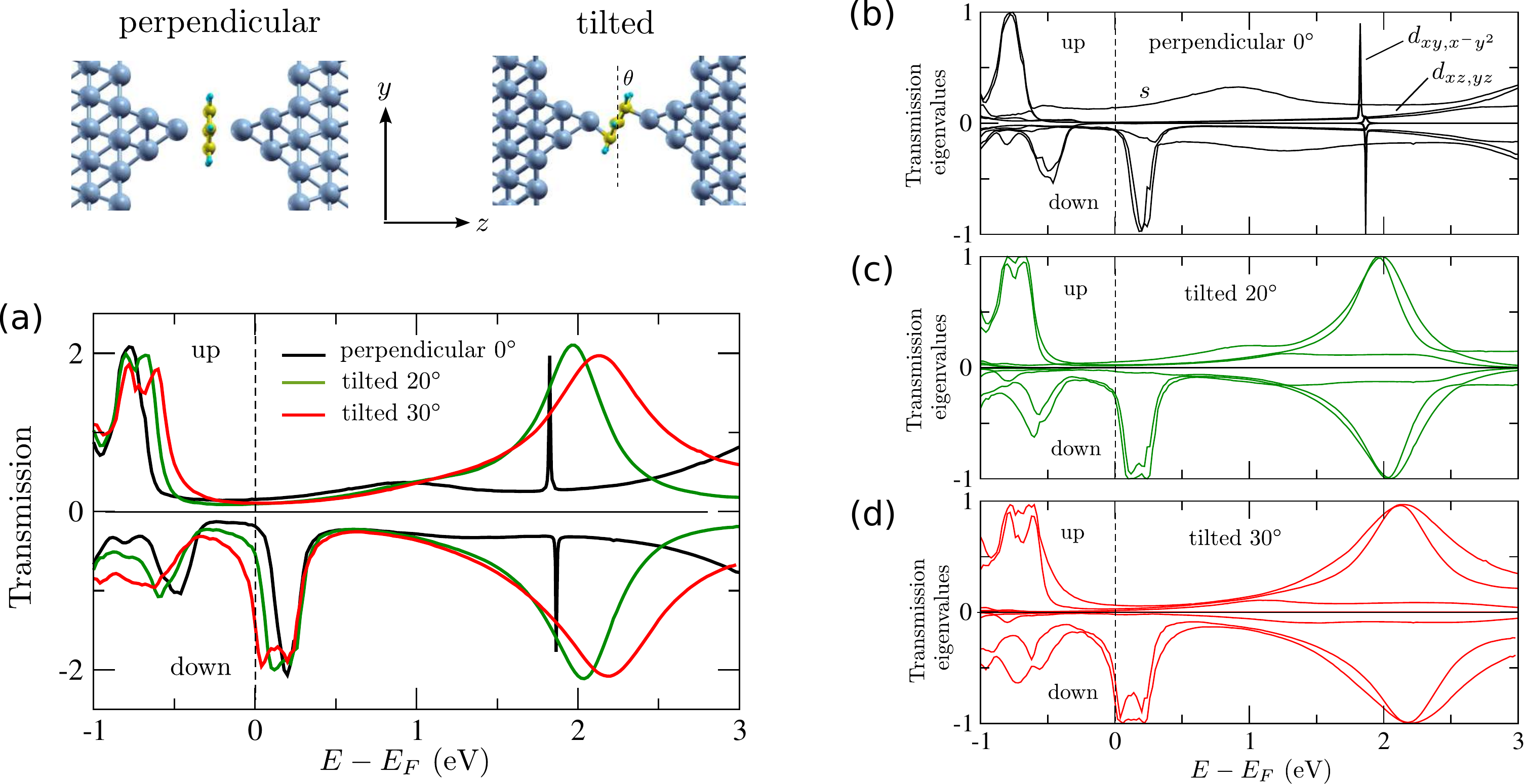}
	\caption{
		TB calculations for Benzene junctions with realistic fcc-Ni(111) electrodes: 
		(a) spin-resolved transmissions as a function of tilting angle $\theta$;
		(b-d) transmission eigenchannels for perpendicular and tilted configurations with $\theta=20^{\circ}$ and $30^{\circ}$.
	}
	\label{Nielectrode}
\end{figure*}

\subsection{Interference from scattering phase shifts}

The actual shape of spin up transmission around the Fermi energy resulting from interference effects mentioned above 
can be well rationalized in terms of so-called scattering phase shifts (see, e.g. Ref. \cite{Baruselli2013}). 
We illustrate the idea on a simple three-level TB model, as shown in Fig.~\ref{phaseshifts},
with the Hamiltonian:
\begin{equation}
\begin{aligned}
\sum_{\alpha \in L,R} [\varepsilon_0{\hat c}_{\alpha}^\dagger {\hat c}_\alpha+
(t{\hat c}_{\alpha}^\dagger {\hat c}_{\alpha+1}+h.c.)]+ \quad\quad\quad\quad\quad \\
\sum_{i=1}^3 [ \varepsilon_i{\hat c}_i^\dagger {\hat c}_i + 
(t_{iL} {\hat c}_i^\dagger{\hat c}_{L0}+t_{iR}{\hat c}_i^\dagger{\hat c}_{R0} + h.c.)],
\end{aligned}
\end{equation}
where the first term stands for the left and right semi-infinite chains with a single band
simulating the spin up $s$-band of a Ni chain. The second term describes the three levels coupled 
to apex atoms (numbered as 0) of the chains. Here, the first level is placed at -7 
(in arbitrary energy units) simulating the important HOMO-2 
benzene orbital contributing to a tunneling transmission at the Fermi energy (set to 0).
Two other levels, set at $-2$ and $2$, simulate HOMO1 and LUMO1 orbitals, respectively, 
which could couple to the chains in the tilted benzene configuration. 
For this single band case the transmission probability can be expressed\cite{Baruselli2013} as 
$T=\sin^2(\delta_e-\delta_o)$, where $\delta_{e/o}$ are phase shifts in
even/odd combination of two electron waves coming from the left and from the right 
chains caused by a coupling to the molecule (levels). 
The reference system (zero phase shifts) is the one when the two chains are fully disconnected (no molecule). 
From the other side, by Friedel sum rule, $d\delta_{e/o}(E)/dE=\pi \Delta\rho_{e/o}(E)$, where $\Delta\rho_{e/o}$ are additional DOS of 
even or odd symmetry (with respect to the transport direction) due to the molecule.

Figs.~\ref{phaseshifts}(a-c) represent the case where only level 1 is coupled to the chains
which would correspond to the perpendicular Benzene orientation. Due to the $\pi$-character of HOMO-2 
(which is also true for HOMOs and LUMOs, see SI Appendix, Fig. S1) its coupling to the left/right chains is antisymmetric 
leading to a small (odd) DOS around the Fermi energy, Figs.~\ref{phaseshifts}a. 
That generates rather constant $\delta_o$ (Fig.~\ref{phaseshifts}b)
which results in a small tunneling transmission around zero energy, as seen in Fig.~\ref{phaseshifts}c.
The case of tilted orientation is simulated in Figs.~\ref{phaseshifts}(d-f) when two other levels are also included. 
Their coupling to the chains should be symmetric (level 2) and antisymmetric (level 3) as can be clearly inferred from 
HOMO1 and LUMO1 orbitals shown in Fig.~\ref{Ni1Dchain}c.  
That will first increase rapidly $\delta_e$ and then $\delta_o$ turning twice the transmission exactly 
zero on both sides of the Fermi energy where $\delta_o=\delta_e$ which agrees clearly with our 
previous calculations, Fig.~\ref{Ni1Dchain}a.  
Very interestingly, the efficient suppression of transmission around the Fermi energy is only achieved if the levels 2 and 3 (HOMO1 and LUMO1)
have different even/odd symmetries, like in the discussed case. If they had the same symmetry, as it is presented in Figs.~\ref{phaseshifts}(g-i) 
for symmetric couplings, the transmission would drop to zero always to the right of either HOMO1 or LUMO1 peaks remaining finite in between.
Note that, unlike transmissions, the molecular DOS shows very similar (trivial) shape for both types of couplings,  Figs.~\ref{phaseshifts}d,g --
two Lorentzian-like peaks from levels 2 (HOMO1) and 3 (LUMO1) superposed on a smooth feature from the level 1 (HOMO-2).

\begin{figure*}[t]
	\centering
	\includegraphics[width=1.0\linewidth]{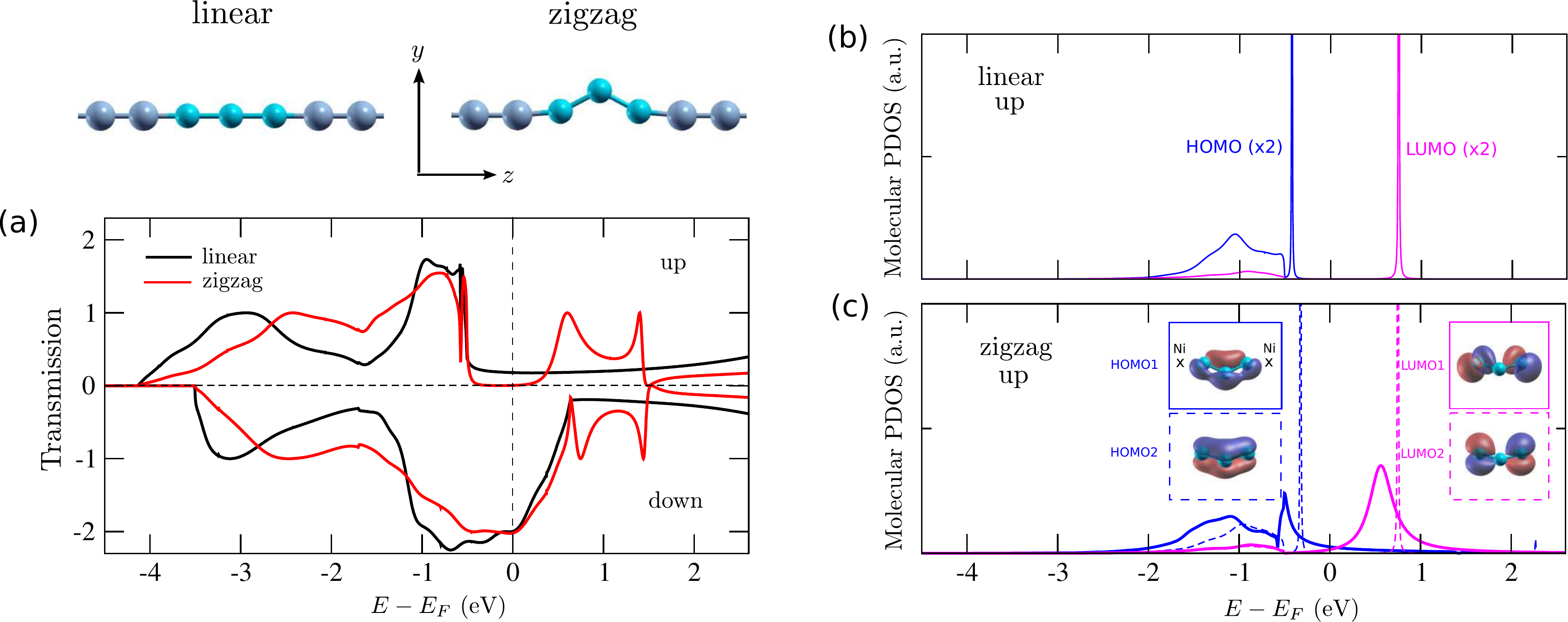}
	\caption{
		TB transport calculations with model Ni chain electrodes for
		a 3-atom Si chain in linear and zigzag geometry: 
		(a) spin-resolved transmissions for two Si chain geometries.
		Due to destructive interference from HOMO1 and LUMO1 orbitals, turned on in zigzag geometry, 
		the spin up transmission gets fully suppressed at the Fermi energy; 
		(b) PDOS on the Si chain for spin up channel in linear configuration; 
		(c) the same for zigzag configuration. Wavefunctions of all the HOMO and LUMO orbitals are shown on insets.
		Only HOMO1 (solid blue) and LUMO1 (solid pink) orbitals, due to their symmetry (even with respect to the $YZ$ plane), 
		could overlap with a Ni $s$-channel.
	}
	\label{Si1Dchain}
\end{figure*}

\subsection{Ni(111)/benzene junctions}

Having understood the main idea on simple model systems we pass now to more realistic nanojunctions by replacing semi-infinite Ni wires with 
fcc-Ni(111) electrodes as it is shown in Fig. \ref{Nielectrode}. Here we show three different geometries as in the case of model systems discussed above. Fig. \ref{Nielectrode}a presents the spin-resolved transmission for those configurations. 
Clearly, spin down conductance is again significantly enhanced when the molecule is tilted
due to broadening of transmission peak centered at around $E=0.2$~eV.
This peak is due to the offset of extra $d_{xz,yz}$ Ni states
and its broadening reflects stronger hybridization with molecular states (mainly with HOMO) 
in low symmetry tilted configuration -- similarly to the case of model Ni chains.

The situation for spin up polarization is more complex: the conductance is only slightly reduced in tilted geometries
and the transmission curve does not show now clear signatures of interference (points of zero transmission).
In order to get a better understanding we plot in Fig. \ref{Nielectrode}b-d spin up transmission eigenvalues for 
all configurations. 
As discussed in the previous section, spin up transmission with Ni chains was essentially provided by a single $s$-channel in a broad energy 
window [-0.5 eV; 2.7 eV]. On the contrary, as can be learned from perpendicular configuration in Fig. \ref{Nielectrode}b, 
five conduction channels are available for transport with realistic Ni electrodes. Those channels 
are mainly provided by $s$, $d_{xz,yz}$, and $d_{xy,x^2-y^2}$ orbitals of Ni apex atoms so we label them 
accordingly in Fig. \ref{Nielectrode}b.   
Two LUMO orbitals can only couple by symmetry (see Appendix A, Fig. \ref{free-bz}) to $d_{xy,x^2-y^2}$ channels
and give rise to sharp (because of weak coupling) peaks in transmission at about 1.8 eV.
The other three channels are orthogonal to the LUMO orbitals and are therefore in off-resonant regime.
Rather constant eigenvalue, of about 0.2 around the Fermi energy, is provided by the $s$ channel 
which clearly dominates in the large energy window. Note that its transmission at the Fermi energy is slightly
lower then the one for model Ni chain case, about $0.35$ (see Fig.~\ref{Ni1Dchain}a), reflecting the smaller $s$-orbital DOS
at the Ni apex atoms (not shown). Finally, two degenerate $d_{xz,yz}$ channels show small transmission of about 0.05 at the LUMO position slowly increasing with energy.

When the molecule is tilted, due to symmetry reduction, the LUMO1 (symmetric with respect to the $YZ$ plane) starts to hybridize with 
the (symmetric) $s$ channel but also, and much stronger, with the (symmetric) $d_{yz}$ one as can be clearly seen for the tilting angle 
of $20^{\circ}$ (Fig. \ref{Nielectrode}c). 
That results in a much more pronounced broadening of the LUMO1 orbital compared to the model Ni chain case. 
Moreover, two anti-crossing points about 1.3 eV and 2.3 eV appear between the two symmetric channels.
Interestingly, the transmission of $s$ channel at the Fermi energy is again suppressed with respect to the one in perpendicular configuration.
In addition, the (antisymmetric) LUMO2 orbital couples strongly to the (antisymmetric) $d_{xz}$ channel giving rise to another broad transmission feature at around 2 eV.
We note that two other channels, of $d_{xy,x^2-y^2}$ symmetry, do not almost contribute to the transport at the energies of interest,
from $-0.5$ to $3$ eV. When the molecule is further tilted, the hybridization of both LUMO orbitals (mainly to $d_{xz,yz}$ Ni states) is increased 
as can be seen for the tilting angle of $30^{\circ}$ (Fig. \ref{Nielectrode}d). 
Finally, we note that in all discussed cases (Fig.~\ref{Ni1Dchain}, \ref{Nielectrode}) the electrode/electrode separation was kept fixed, 
$D=4.39$~\AA, in order to analize clearly the effect of molecule tilting.
The effect of increasing $D$ on spin polarization was found to be much smaller 
leading to a trivial decrease of both spin up and down conductances (see SI Appendix, Fig. S2).   
We can conclude therefore that by tilting the molecule the $s$ channel
contribution to the spin up conductance is suppressed due to quantum interference but the overall
effect is somewhat hindered because of additional $d$-channels appearing in the case of realistic Ni electrodes.

\subsection{Three-atom Si junctions with Ni chain electrodes}

Discussed so far symmetry arguments to control spin-filtering property are in fact quite general and can be realized in some other situations.  
We have found, for example, that similar interference patterns in transmission function can appear for some atomic chains (see Fig. \ref{Si1Dchain}, upper panel) during transition from linear to zigzag geometry. Such chains can be realized in many possible ways: for example, out of various kinds of alkanes or alkenes molecules \cite{Su-2016}. To demonstrate once again our principle we have chosen a simple model 3-atom Si chain which could be stretched from a zigzag geometry (energetically more preferable) to almost linear configuration before being broken. 
In the linear configuration, 2-fold degenerate HOMO and LUMO (originated from $p_x$ and $p_y$ Si orbitals) are again strictly orthogonal by symmetry to the $s$-like Ni channel and are therefore fully transparent for electron transport around the Fermi energy which 
is dominated by a tunneling through other $p_z$-character Si orbitals (not shown in the figure).
In zigzag configuration, due to lack of full axial symmetry, the HOMO1 and LUMO1 orbitals start to overlap
and to interfere destructively with the background tunneling $s$-channel (all states being symmetric in the $YZ$ mirror plane). 
The HOMO2 and LUMO2, being antisymmetric, remain orthogonal to the $s$-channel and appear as $\delta$-like peaks in molecular DOS at about -0.3 eV and 0.8 eV, respectively. This results again in complete suppression of spin up transmission around the Fermi energy and to the full spin polarization of electric current in zigzag geometry. Note that the HOMO1 and LUMO1 have again different even/odd symmetry
with respect to the $XY$ plane perpendicular to the transport direction and passing through the central Si atom -- the 
important condition to suppress efficiently the transmission in the HOMO-LUMO gap region.

\section{Conclusions}
We have suggested a possible way to manipulate the spin-filtering property of single molecule junctions 
based on clear symmetry arguments. The important point is the change of nanojunction symmetry -- by, for example, a mechanical strain --
which can switch on/off new channels for electron propagation (via molecular orbitals which were before inactive)
and thus induce new interference patterns in electron transmission function around the Fermi energy.
On example of two representative systems, benzene molecule and 3-atom Si chains bridging two Ni electrodes,
we have shown that when the symmetry is lowered -- tilted benzene molecule or zigzag Si chain --
spin up conductance gets strongly suppressed due to destructive interference of main $s$-channel with (frontier) HOMO and LUMO orbitals.
It was demonstrated moreover that those orbitals must have different (even/odd) symmetry with respect to the transport direction
in order to provide strong suppression of spin up transmission, which is the case of both molecular junctions.
Spin down conductance is, on the contrary, enhanced due to stronger molecule-metal hybridization. Suggested symmetry arguments were shown to work perfectly for model systems with Ni chain electrodes, giving fully spin-polarized conductance 
in low symmetry configurations, and slightly worse for realistic Ni electrodes due to multichannel character of electron 
transport around the Fermi level. 
Finally, it should be noted that the ratio of spin-filtering modulation depends on a background value of ($s$-channel) spin up conductance
(to be suppressed) which was rather small in our case but could be probably found much larger for some other systems.       
We believe that our findings will be of importance not only from conceptual point of view but also for  
exploring novel functionalities for spin-based devices involving solid symmetry arguments and quantum
interference effects.

{\bf Acknowledgments}: This work has been performed using HPC computation resources from TGCC-GENCI (Grant No. A0040910407). 
D.L. was supported by the Alexander von Humboldt Foundation through a Fellowship for Postdoctoral Researchers.
\\

\section*{Appendix A: Electronic structure of free Benzene molecule}

\begin{figure*}
	\includegraphics[scale=0.7]{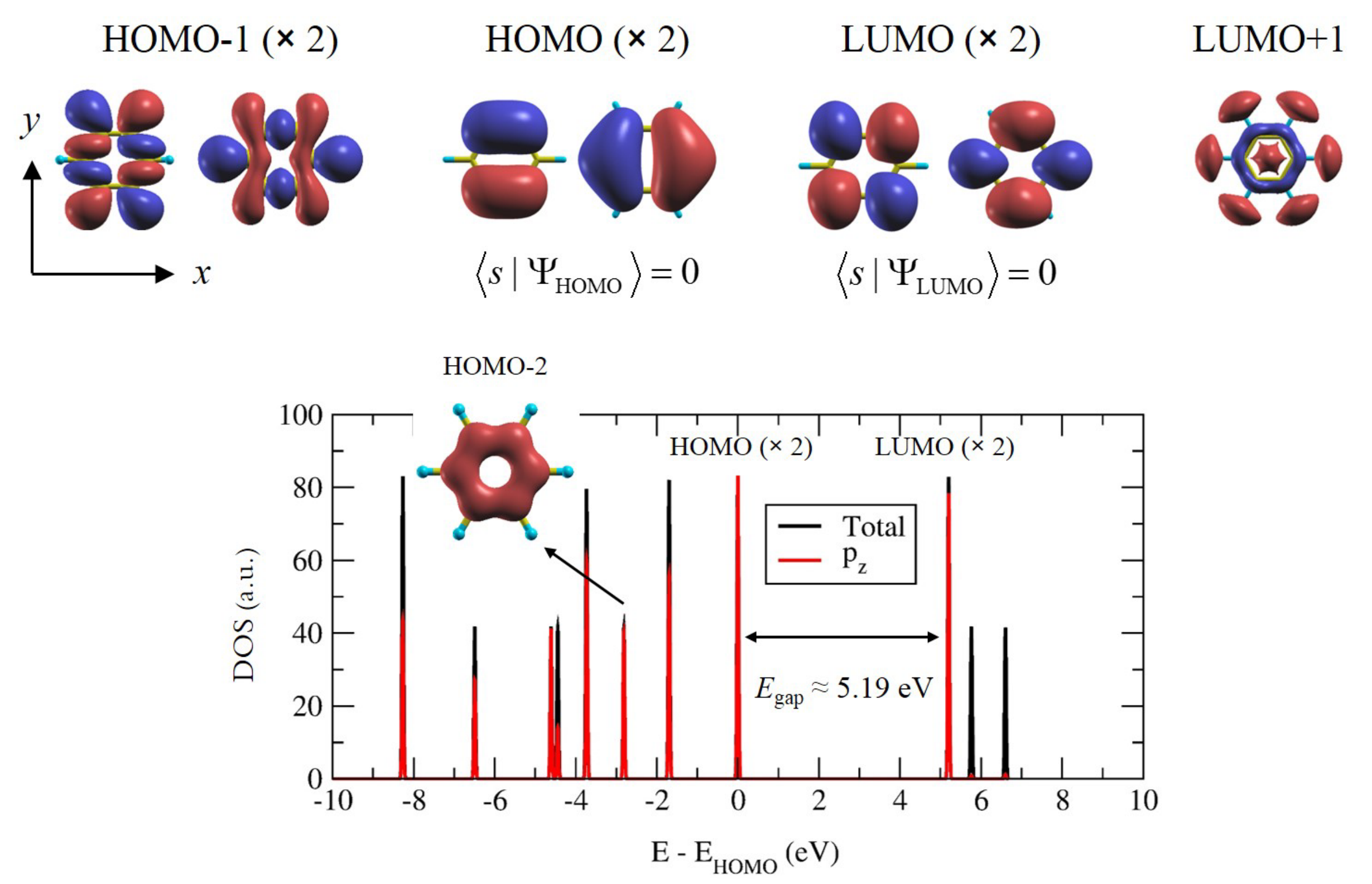}
	\caption{Density of states (DOS) and projected DOS on Carbon $p_z$ orbitals for a free benzene molecule by DFT calculations. We present in addition the spatial distributions of some orbital wavefunctions. Isosurfaces of positive and negative isovalues are shown in red and blue, respectively. \label{free-bz}}
\end{figure*}

In Fig.~\ref{free-bz}, we plot the density of states (DOS) of benzene molecule in gas phase and wave functions of some important molecular orbitals,
in particular, HOMO-2, HOMO, and LUMO. The calculated HOMO-LUMO gap at the DFT level was found about 5.19 eV. By orbital symmetry, the orbitals around the Fermi energy, including 2-fold degenerate HOMO and LUMO, are found all to be of odd symmetry (with respect to molecular $XY$ plane) being $\pi$-states originating mainly from Carbon $p_z$ atomic orbitals. In addition, only HOMO-2 (bonding) orbital would have nonzero 
overlap with the Ni s-orbitals when the Benzene molecule is placed in between two Ni chains in perpendicular orientation, 
providing the tunneling transmission at the Fermi level. The HOMO and LUMO orbitals can overlap with only $d_{xz,yz}$ and $d_{x^2-y^2,xy}$ Ni orbitals, respectively. For tilted Benzene configuration, as discussed in the main text, due to symmetry reduction also one of HOMO and LUMO orbitals will start to overlap with the Ni $s$-channel.  

\section*{Appendix B: The effect of electrode/electrode separation on spin-polarized conductance}
\begin{figure*}
	\includegraphics[scale=0.65]{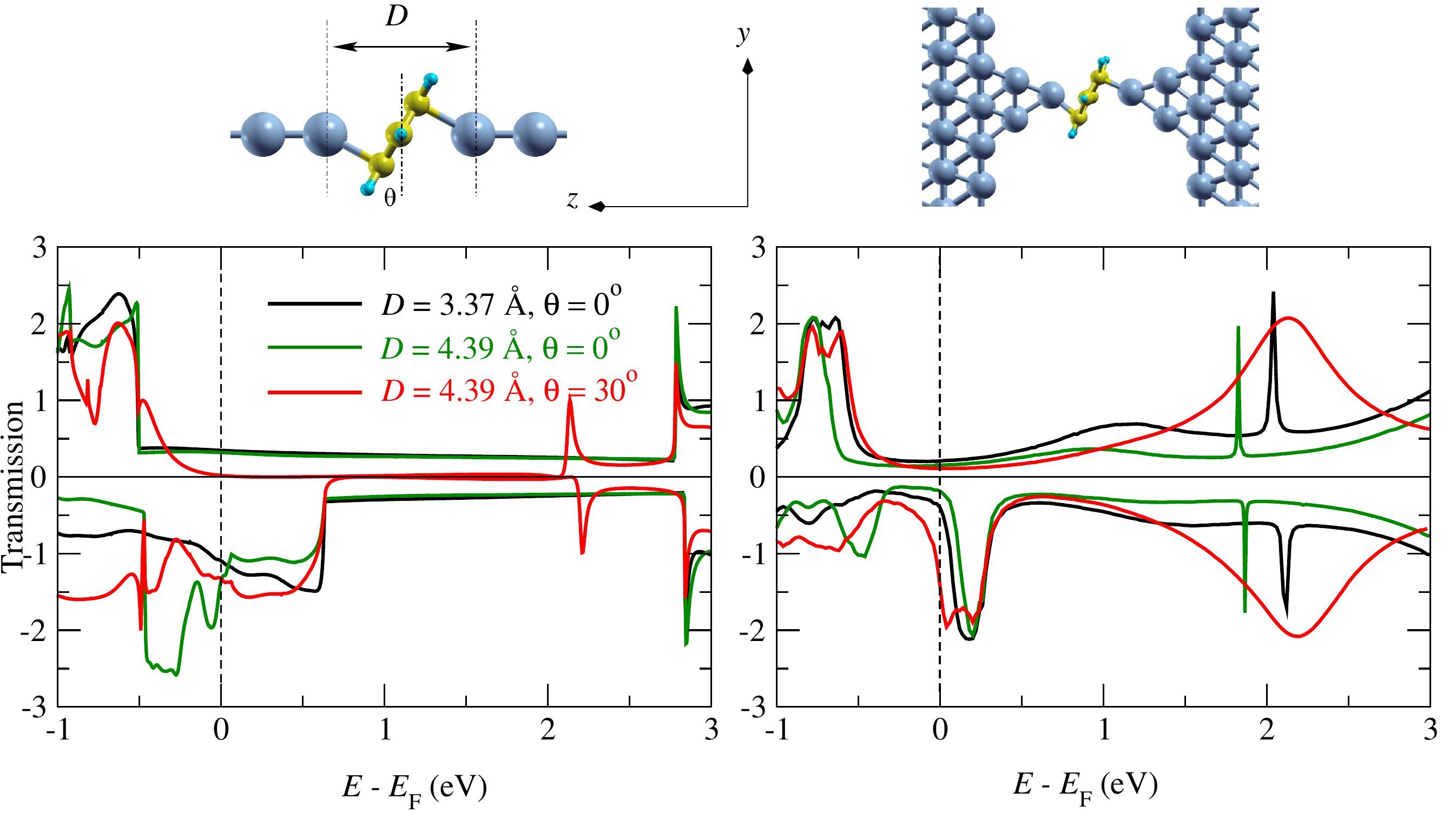}
	\caption{\label{sup2} TB results for spin-resolved transmission functions for different tilting angle $\theta$ and electrode/electrode separation $D$. The results for both model Ni chains (left) and realistic Ni(111) electrodes (right) are presented.}
\end{figure*}

The main goal of the paper was to analize in detail the effect of Benzene tilting on spin-polarized transmission as a consequence of symmetry reduction. Therefore the electrode/electrode separation was everywhere fixed to 4.39 \AA corresponding to the one where double-bond configuration (tilted Benzene) is realized as shown in Fig.~1. In reality however the perpendicular configuration, at tilting angle $\theta=0$, is realized at smaller 
$D$. Fig.~\ref{sup2} demonstrates however that the shorter $D$, from 4.39 to 3.37~\AA, leads to rather trivial small increase of both 
spin up and down conductances for both model and realistic Ni electrodes, contrary to much more pronounced effect of Benzene tilting 
affecting quite differently two spin channels.

\bibliographystyle{apsrev}
\bibliography{manuscript}

\begin{thebibliography}{34}
\bibitem[{com()}]{comment2}
\bibinfo{note}{The spin polarization is defined as
  ${SP}=(G_{\uparrow}-G_{\downarrow})/(G_{\uparrow}+G_{\downarrow})$, where
  $G_{\uparrow}$ and $G_{\downarrow}$ are the conductances of majority and
  minority spin channels, respectively.}
\end{thebibliography}

\end{document}